\documentstyle[12pt,axodraw]{article}
\title{Single $W$-boson Production at Linear Colliders
         }
\author{
         E.E.~Boos, M.N.~Dubinin \\
     {\small \it Institute of Nuclear Physics, Moscow State University} \\
       {\small \it  119899 Moscow, Russia} \\}
\date{}

\begin{document}
\maketitle

\begin{abstract}
The process of single $W$ boson production at the energies of
Next Linear Colliders is considered. We discuss in details the
contributions from $s$ and $t$ channel diagrams, technical
aspects of the complete tree level calculation with a finite
$W$ width, and the quark mass effects.
\end{abstract}

The process of single $W$-boson production $e^+ e^- \to e^- \bar \nu_e
u \bar d$ ("CC20 process" in the four fermion channels classification
of \cite{eventWW}) has been considered in details both from experimental
and theoretical viewpoints. From the experimental point of view this
process gives an important contribution to the cross section of $W^+ W^-$
pair production as well as to the rate of single $W$ production, when
the initial electron (positron) goes to the beam pipe. At the same
time CC20 process
represents the main background to signals of new
physics (especially $R$-parity conserving or violating SUSY processes)
and provides a strong restrictions on the anomalous three vector
boson couplings $WW\gamma$, $WWZ$. From the theoretical point of view 
the cross section of $e^+ e^- \to e^- \bar \nu_e u \bar d$ channel
is sensitive to a huge gauge cancellations between the individual
diagrams from the complete tree level set, and the singularity of
the amplitude at zero scattering angle of the electron, when the
treatment of finite $W$-boson width and finite electron mass 
is delicate and needs much care \cite{theor1,theor2}. It will 
be also pointed out that the ISR corrections to CC20 process should be
calculated with different characteristic energy scales 
for the two gauge invariant subsets of contributing diagrams.

The set of 20 diagrams of the process CC20 can be divided into two gauge
invariant (with respect to the SM gauge group) subsets of 10 $t$-channel
(Fig.1) and 10 $s$-channel diagrams (Fig.2).
If the lepton and the corresponding neutrino in the final state of
$e^+ e^- \to e^- \bar \nu_e u \bar d$
are replaced by the muon or tau with corresponding neutrino, only 10
$s$-channel diagrams remain and represent the analogous $\mu$ and
$\tau$ CC10 channels.
Electron scattering at zero angle from the $t$-channel subset is absent
in these channels. It is a simple proof that both CC10 channels are
gauge invariant. The less trivial statement is that these two CC10 
classes are the minimal gauge invariant SM subsets. In fact it follows
from the general theorem proved recently in \cite{boos_ohl}.

Our calculation of the total rate have been performed by means of
{\it Comp\-HEP} package \cite{CompHEP}. The squared amplitude with finite
electron
mass $m_e=$ 0.511 MeV have been used in this calculation, so the forward
electron pole is regulated by the kinematical cutoff $t_{max}=-m_e^2
(M^2/s)^2$, where $M$ denotes an invariant mass scale for the $\bar \nu_e
u \bar d$ system (see Fig.3). 
Results of $CompHEP$ calculation are shown in Table 1
together with the results of other groups obtained by means of $grc4f$
\cite{grace},
$KORALW$ \cite{koralw} and $WPHACT$ \cite{wphact} generators. $grc4f$
and $WPHACT$ generators also use the finite fermion mass amplitude and
$KORALW$ contains $grc4f$ matrix element inside. Their phase space
intergation routines are different. Good agreement between the three
generators ($CompHEP$, $grc4f$, $WPHACT$) is observed while $KORALW$
results obtained for $\sqrt{s}=$500 GeV are somewhat smaller.
The agreement of cross sections is not trivial since
$CompHEP$, $grc4f$ and $WPHACT$ are using different prescriptions
for the insertion of Breit-Wigner propagator into the complete tree
level amplitude.

In the 'overall' prescription (also called the 
'preserved gauge' scheme) used by $CompHEP$ and $WPHACT$, resonant and
nonresonant parts of the amplitude
\begin{equation}
\frac{a_{\mu}}{p^2_W-m^2_W} + b_{\mu}
\end{equation}
are squared, summed and then multiplied by the 'overall' factor
\begin{equation}
\frac{(p^2_W - m^2_W)^2}{(p^2_W - m^2_W)^2 +  m^2_W \Gamma^2_W}
\end{equation} 
In $CompHEP$ the number of overall factors is equal to the number 
of $W$ propagators in the diagram set, and the prescription has beed
applied to two gauge invariant CC10 subsets separately in order
to avoid an artificial suppression of CC10-t part in the phase
space region close to the position of the second $W$ pole from the CC10-s
part.

$grc4f$ and $KORALW$
prescriptions \cite{theor1} are based on the gauge-invariance motivated 
redefinition of the leptonic tensor 
$\L_{\mu \nu}$ \cite{theor1,gutbrod} for the $t$-channel subset
\begin{eqnarray*}
L_{\mu \nu}&=
  &2(p_{\mu} p^\prime_{\nu} + p_{\nu} p^\prime_{\mu}) + q^2 \,g_{\mu \nu}
\end{eqnarray*}
to the form
\begin{eqnarray*}
L^\prime_{\mu \nu}&=&4(p_{\mu} -\frac{p_0}{q_0}\, q_{\mu})
               (p_{\nu} -\frac{p_0}{q_0}\, q_{\nu})  + q^2 \, g_{\mu \nu}
\end{eqnarray*}   
Unadequate treatment of finite $W$ width violates the gauge
cancellation of the double pole $1/t^2=1/(p^\prime_e-p_e)^4$ for
$t$-channel gamma and
leads to nonunitary (powerlike) energy behaviour of the amplitude. 
The overall prescription or the redifinition $L_{\mu \nu} \to L^\prime_{\mu
\nu}$ ensures a numerically stable can\-cel\-la\-tion of the double pole
to the single one $1/t$. This cancellation in the preserved gauge scheme 
used in $CompHEP$ can be explicitly demonstrated if
we plot the distribution in $log_{10} \, |t|$ (see Fig.3 (1)). Flat part
of the distribution is related to the unitary behaviour $d\sigma/dt \sim
1/t$. In the framework of the equivalent gamma approach this
behaviour corresponds to the canonical Wieszacker-Williams approximation.
The falldown of the distribution starting at about $log_{10} \, |t| \sim$
-6-7 reflects an improved Wieszacker-Williams behaviour  \cite{mangano}
The $W$ peak is observed in Fig.2 at $log_{10} \, |t| \sim$ 4.
\footnote{Detailed comparison of the same kind as indicated above at LEP2
energies
$\sqrt{s}=$ 183 and 190 GeV can be found in \cite{ballestrero}.}

We present the results of $CompHEP$ calculation for contributions
of separately taken $s$-channel and $t$-channel subsets to the total rate,
as well as
their interference contribution, at various energies in Table 2. Quark
phase space cuts $E_q > 3$ GeV, $M_{ud} > 5$ GeV and/or lepton phase space
cut $\cos\vartheta_e >$ 0.997 are imposed. 
From Table 2 and Fig.4 one can see that the contribution of
$t$-channel   
subset increases rapidly with energy while the $s$-channel cross section
goes down and becomes smaller than the $t$-channel at the energy
about 320 GeV. The $s$-$t$ interference is extremely small in all
the energy range under consideration (100 GeV - 1 TeV). At LEP2 energies
the single $W$ production process is less important than the $W^+W^-$ pair
production,
but at the energies of LC single $W$ creation plays the dominant role.

Table 3 contains the cross
section values at different energies without cuts, when the $t$-channel
gamma pole and the $t$-channel $u/d$ quark poles in the ladder diagram
are regulated by the electron and light quark masses $m_u=$ 5 MeV, $m_d=$
10 MeV. It is important to point out that at nonzero electron mass the
finite numerical result exists even for massless quarks. If we denote
the minimal quark momentum fraction by $x_{min}$
\begin{eqnarray*}
x_{min}=\frac{(m_u+m_d+m_e)^2}{s}
\end{eqnarray*}
the maximal gamma momentum transferred
\begin{eqnarray*}
t_{max}= - m^2_e \, \frac{x^2_{min}}{1-x_{min}}
\end{eqnarray*}
and the kinematical cutoff near the pole still exists at $m_u=m_d=$ 0. 
It is important to understand a role of the multiperipheral
diagrams in the CC10 $t$-channel subset. However, it makes no sense
to compute straightfrwardly  a contribution from that diagrams because
they do not form a gauge invariant class. But one can get an information
about a size of the multiperipheral diagrams contribution 
by calculating a dependence of the cross section on a quark mass.
These diagrams represent the only production mechanism that
could depend significantly on the values of quark masses. We
calculated the cross section dependence on the quark
mass $m_q=m_u=m_d$ changing from 1 MeV to 100 MeV (see Table 4 and Fig.4).   
The total rate decreases from 1442 fb to 1414 fb, so rather weak
but distinct dependence on the quark mass takes place. 
Therefore one can expect a rather small 'resolved photon' contribution to
CC20 process rate. Instructive comparison can be done with the case
of resolved photon contribution to the single $W$ production process at
HERA ($ep$ collisions), where the latter is estimated to be of order
10-15\% of the total $W$ rate \cite{resolved}. In the case of $ep$
collisions the most important $t$-channel diagram topology (see diagram 7
in Fig.1) contains a quark line and $t$-channel gamma, giving rise to
potentially high resolved cross section $\gamma^* q \to$ {\it hadrons}. In
the case of $e^+ e^-$ collisions this contribution is absent and only
subleading multiperipheral topologies (see diagrams 2,3 in Fig.1)  
could give the resolved $\gamma^* q \to$ {\it hadrons}.

The calculation of initial state radiative corrections (ISR) to CC20
process needs a special care. Two different scales of ISR radiator
function should be used for the $t$ and $s$ channel subsets of diagrams.
If we take $Q^2=s$ for $s$ channel and $Q^2=0$ for $t$-channel, the cross
section can be calclated as a sum of $s$-channel contribution with ISR
and $t$-channel contribution without ISR from the Table 2.
The same is also true for the value of $\alpha_{QED}$ which should be 
taken differently for $t$-channel and $s$-channel sets. Obviously, the
characteristic
scale for the $t$-channel part is $Q^2=0$ and therefore one should 
use 1/137 while a typical scale of the $s$-channle piece is of the order
$s$ and a value of about 1/128 should be used.

The authors are grateful to E.~Accomando, A.~Ballestrero and G.~Passarino
for useful discussions and to DESY-Zeuthen TESLA group for the kind
hospitality. This work was partly supported by the joint RFBR-DFG grant
99-02-04011.

\newpage
 
\begin{figure}[hb]
\begin{center} 
\unitlength=1cm
\begin{picture}(16,10)
\put(-5.8,-21){\epsfxsize=24cm \epsfysize=35cm \leavevmode
\epsfbox{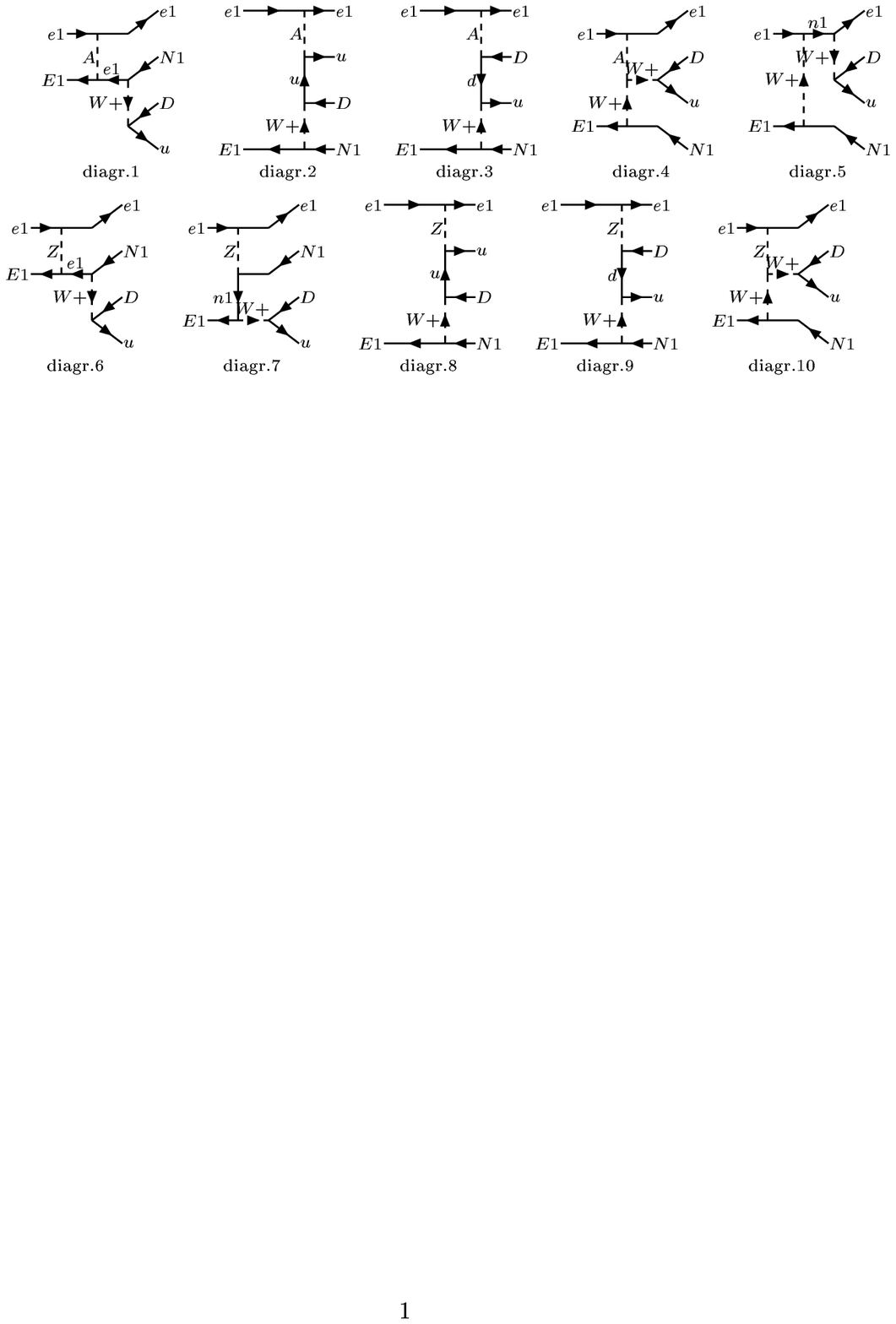}}
\end{picture}
\end{center}
\caption{t-channel Feynman diagrams for the process
$e^+ e^- \rightarrow e^- \bar \nu_e u \bar d$.
}
\end{figure}  

\newpage

\begin{figure}[h]
\begin{center}
\unitlength=1cm
\begin{picture}(16,10)
\put(-5.8,-21){\epsfxsize=24cm \epsfysize=35cm \leavevmode
\epsfbox{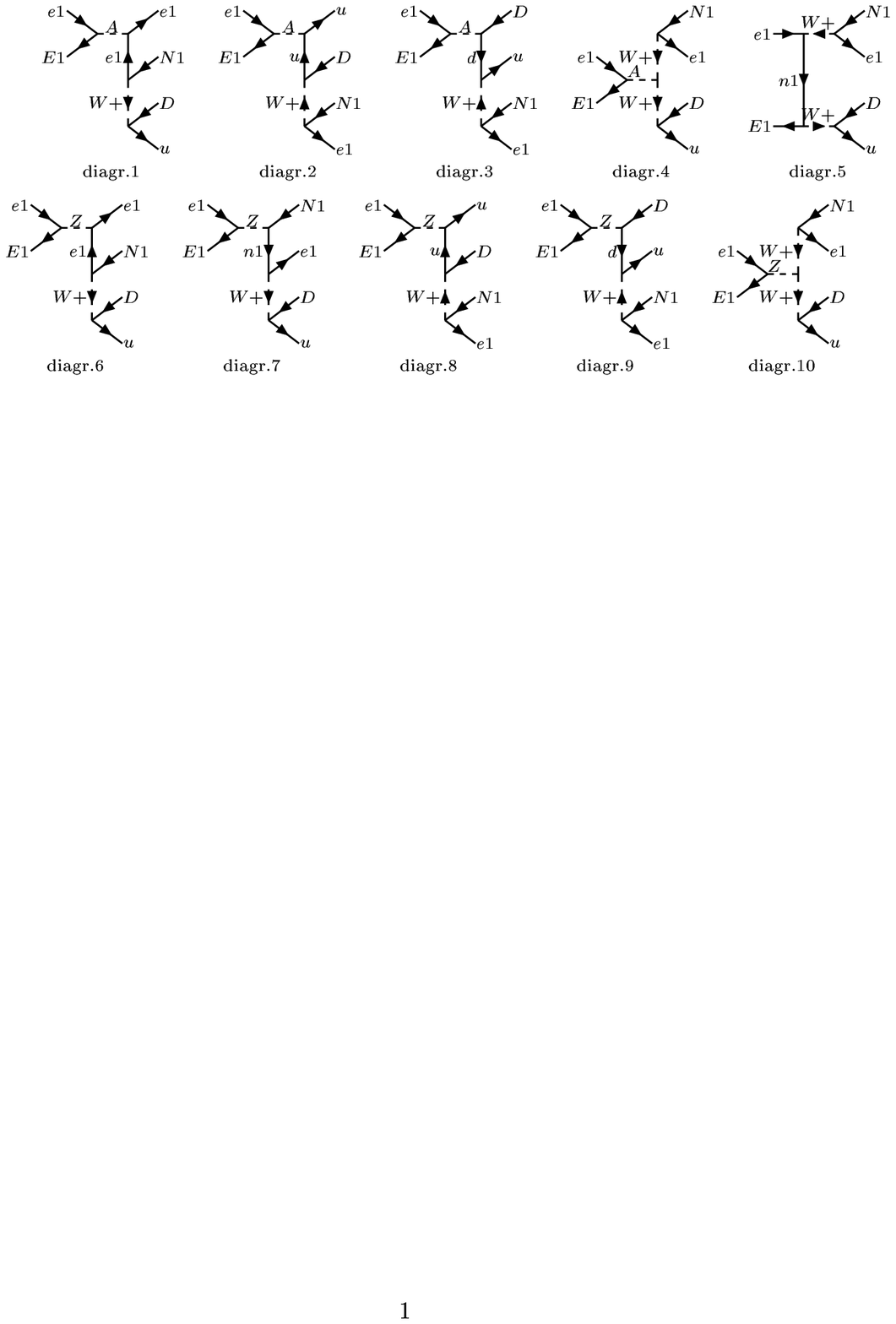}}
\end{picture}
\end{center}
\caption{s-channel Feynman diagrams for the process
$e^+ e^- \rightarrow e^- \bar \nu_e u \bar d$.
}
\end{figure}

\newpage

\begin{table}[t]
\begin{center}
\begin{tabular}{|c|c|c|c|c|}
\hline
    & CompHEP & grc4f & KORALW & WPHACT \\ \hline
\multicolumn{5}{|c|}{$\sqrt{s}=$ 350 GeV} \\ \hline
$q$ cuts, no ISR & 1076(1) & 1080(2) & & 1074(1)  \\  
$q$ cuts, with ISR & 1039(1) & 1040(1) & & 1038(2) \\
$q$, $e$ cuts, no ISR & 520(1) & 521(1) & & 512(1) \\  
$q$, $e$ cuts, with ISR & 478(1) & 480(1) & & 479(2) \\ \hline
\multicolumn{5}{|c|}{$\sqrt{s}=$ 500 GeV} \\ \hline
$q$ cuts, no ISR & 1417(2) & 1419(2) & 1395(6) & 1418(1)\\
$q$ cuts, with ISR & 1357(5) & 1359(2) & 1336(6)& 1358(2) \\
$q$, $e$ cuts, no ISR & 939(3) & 939(1) & 909(5) & 936(1)\\
$q$, $e$ cuts, with ISR & 864(9) & 874(1) & 840(5)& 847(2) \\ \hline
\multicolumn{5}{|c|}{$\sqrt{s}=$ 800 GeV }\\ \hline
$q$ cuts, no ISR & 2140(4) & 2146(3) & & 2138(3) \\    
$q$ cuts, with ISR & 2048(4) & 2046(3) & & 2042(2) \\  
$q$, $e$ cuts, no ISR & 1687(4) & 1697(2) & &1692(3)\\  
$q$, $e$ cuts, with ISR & 1597(3) & 1597(2) & & 1593(2) \\
\hline
\end{tabular}
\end{center}
\caption{CompHEP, grc4f, KORALW and WPHACT results for the total cross
section of the process $e^+ e^- \to e^- \bar \nu_e u \bar d$ (fb) at
$\sqrt{s}=$ 350, 500 and 800 GeV. The notation '$q$ cuts' means partonic
level quark cuts $E_q \ge$ 3 GeV, $M_{ud} \ge$ 5 GeV, '$l$ cuts'
corresponds to $\cos \theta_e \ge$ 0.997 One standard deviation error of
the last digit is indicated in brackets.}
\end{table}

\newpage

\begin{table}[t]
\begin{center}
\begin{tabular}{|c|c|c|c|c|}
\hline
$\sqrt{s}$ & $\sigma(CC10-t)$ & $\sigma(CC10-s)$ & $\sigma(t-s \; 
interf.)$ & $\sigma_{tot}$ \\ \hline
\multicolumn{5}{|c|}{quark phase space cuts, no ISR} \\ \hline
183      & 130(0)         & 655(1)         & 0.2(0)  & 785(1) \\
190      & 147(0)         & 680(1)         & 5(0)    & 832(1) \\
350      & 635(1)         & 420(1)         & 21(0)   & 1076(1) \\ 
500      & 1127(2)        & 270(0)         & 19(0)   & 1417(2) \\
800      & 1981(4)        & 143(0)         & 16(0)   & 2140(4) \\
\hline
\multicolumn{5}{|c|}{quark phase space cuts, with ISR} \\ \hline
183      & 117(0)         & 566(1)        & 0.2(0)   & 683(1) \\
190      & 132(0)         & 603(1)        & 5(0)     & 739(1) \\
350      & 587(1)         & 432(1)        & 20(0)    & 1039(1) \\
500      & 1049(5)        & 289(0)        & 19(0)    & 1357(5) \\ 
850      & 1873(4)        & 159(0)        & 16(1)    & 2048(4) \\
\hline
\multicolumn{5}{|c|}{lepton and quark phase space cuts, no ISR} \\ \hline
183      & 102(0)         & 2(0)          & 0.0(0)   & 104(0) \\
190      & 116(0)         & 2(0)          & 0.0(0)   & 118(0) \\ 
350      & 513(1)         & 7(0)          & 0.2(0)   & 520(1)  \\
500      & 928(2)         & 10(0)         & 0.3(0)   & 938(2)  \\
800      & 1671(4)        & 15(0)         & 0.4(0)   & 1686(4) \\
\hline
\multicolumn{5}{|c|}{lepton and quark phase space cuts, with ISR} \\ \hline
183      & 92(1)          & 2(0)          & 0.0(0)   & 94(1) \\
190      & 103(1)         & 2(0)          & 0.0(0)   & 105(1) \\
350      & 472(1)         & 6(0)          & 0.0(0)   & 478(1) \\
500      & 854(9)         & 10(0)         & 0.3(0)   & 864(9) \\
800      & 1581(3)        & 15(0)         & 0.4(0)   & 1596(3) \\
\hline
\end{tabular}
\end{center}
\caption{ Contributions to the total cross section (in fb) of the CC20
process
$e^+ e^- \to e^- \bar \nu_e u \bar d$ from the gauge invariant subsets
of $t$-channel and $s$-channel diagrams and their interference at the
energies of LEP2 and NLC. One standard deviation error of the last digit
is indicated in brackets.}
\end{table}

\newpage

\begin{table}[t]
\begin{center}
\begin{tabular}{|c|c|c|c|c|}
\hline
$\sqrt{s}$ & $\sigma(CC10-t)$ & $\sigma(CC10-s)$ & $\sigma(t-s \;
interf.)$ & $\sigma_{tot}$ \\ \hline
\multicolumn{5}{|c|}{no cuts, no ISR} \\ \hline
183      & 141(0)         & 655(1)         & 0.2(0)  & 796(1) \\
190      & 158(0)         & 680(1)         & 5(0)    & 843(1) \\
500      & 1141(2)        & 271(0)         & 19(0)   & 1531(2) \\
800      & 2000(6)        & 143(0)         & 16(0)   & 2159(6) \\
\hline
\end{tabular}
\end{center}
\caption{ Contributions to the total cross section (in fb) of the CC20
process
$e^+ e^- \to e^- \bar \nu_e u \bar d$ from the gauge invariant subsets
of $t$-channel and $s$-channel diagrams and their interference at the
energies of LEP2 and NLC. Phase space cuts are not imposed. One standard
deviation error of the last digit 
is indicated in brackets.}
\end{table}

\begin{table}[t]
\begin{center}
\begin{tabular}{|c|c|c|c|c|}
\hline
$m_q$ (MeV) & $\sigma(CC10-t)$ & $\sigma(CC10-s)$ & $\sigma(t-s \;
interf.)$ & $\sigma_{tot}$ \\ \hline
0   & 1160(4) & 270(0) & 20(0) & 1450(4) \\ 
0.5 & 1154(3) & 270(0) & 20(0) & 1444(3) \\
1   & 1152(3) & 270(0) & 20(0) & 1442(3) \\
5   & 1147(2) & 270(0) & 20(0) & 1437(3) \\
10  & 1142(2) & 270(0) & 20(0) & 1432(3) \\
50  & 1131(2) & 270(0) & 20(0) & 1421(2) \\
100 & 1124(2) & 270(0) & 20(0) & 1414(2) \\ \hline
\end{tabular}
\end{center}
\caption{The total rate for the process
$e^+ e^- \to e^- \bar \nu_e u \bar d$ (fb) at various quark masses.} 
\end{table}

\newpage

\begin{figure}[hb]
\begin{center}  
\unitlength=1cm
\begin{picture}(16,10)
\put(-3.0,-1){\epsfxsize=19cm \epsfysize=19cm \leavevmode
\epsfbox{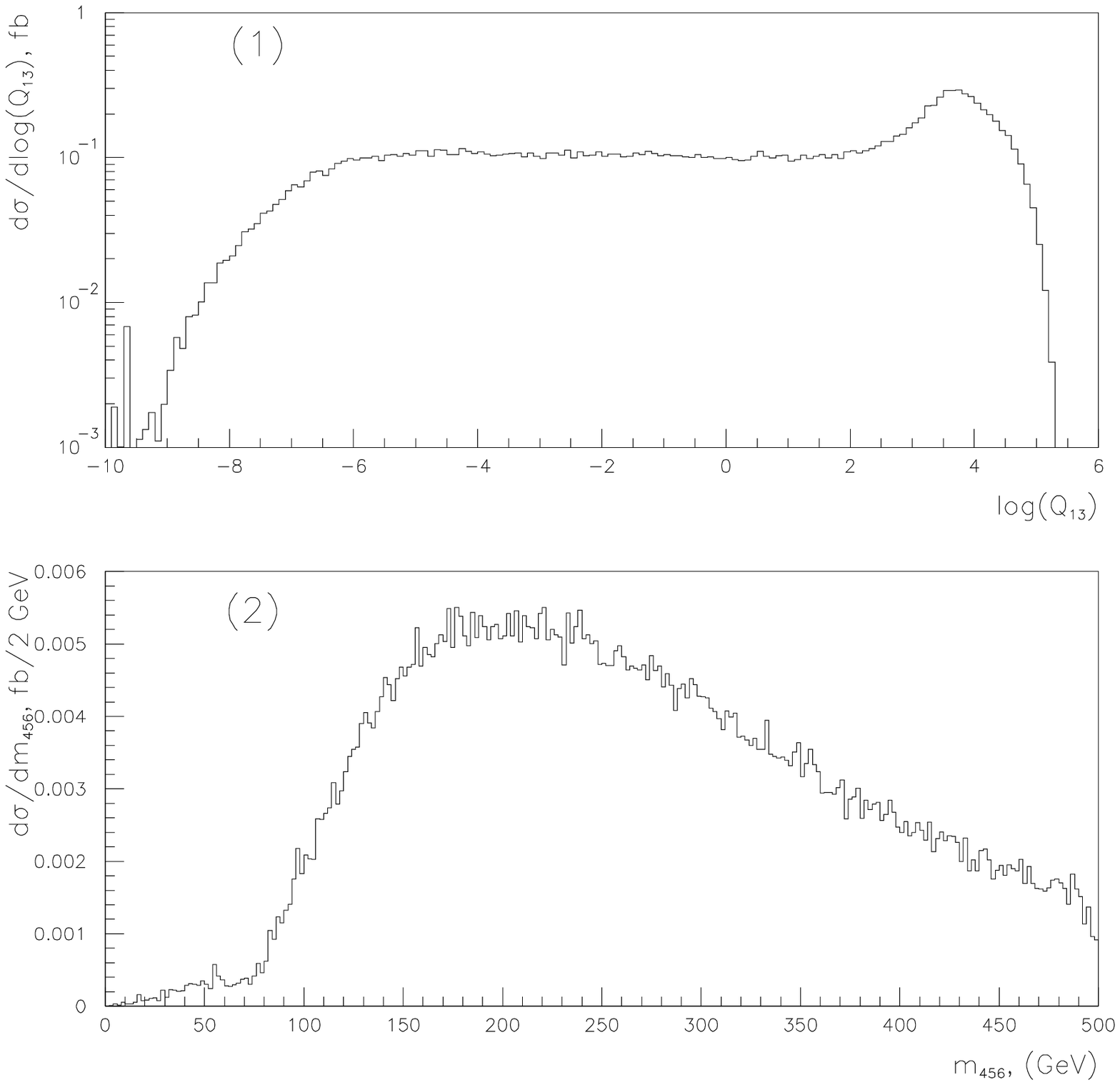}}
\end{picture}
\end{center}
\caption{ (1)- the distribution of $\log_{10} (Q^2)$, 
(2) - the $M_{\nu u \bar d}$ invariant mass distribution }
\end{figure}

\newpage

\begin{figure}[hb]
\begin{center}
\unitlength=1cm
\begin{picture}(16,10)
\put(-1.0,-2.5){\epsfxsize=15cm \epsfysize=20cm \leavevmode
\epsfbox{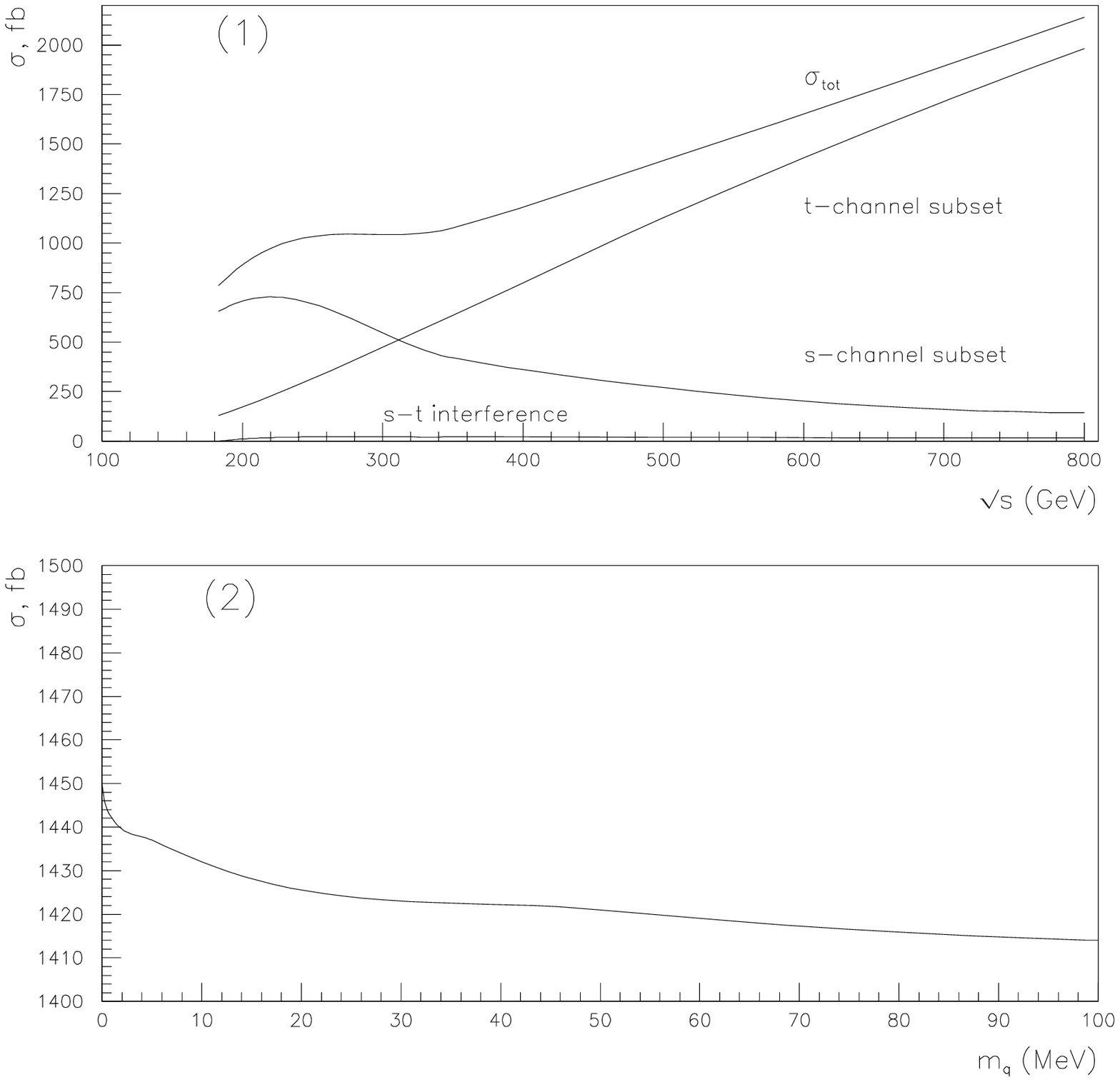}}
\end{picture}
\end{center} 
\caption{ (1)- s and t channel subset contributions to
$\sigma_{tot}$, (2) - total cross section dependence on the quark mass.}
\end{figure}

\end{document}